# Realtime Multimodal Emotion Estimation using Behavioral and Neurophysiological Data

Von Ralph Dane Marquez Herbuela
International Research Center
for Neurointelligence (WPI-IRCN)
The University of Tokyo
Tokyo Japan
herbuela.vonralphdane@mail.u-tokyo.ac.jp

Yukie Nagai
International Research Center
for Neurointelligence (WPI-IRCN)
The University of Tokyo
Tokyo Japan
nagai.yukie@mail.u-tokyo.ac.jp

## ABSTRACT

Many individuals—especially those with autism spectrum disorder (ASD), alexithymia, or other neurodivergent profiles—face challenges in recognizing, expressing, or interpreting emotions. To support more inclusive and personalized emotion technologies, we present a real-time multimodal emotion estimation system that combines neurophysiological EEG, ECG, blood volume pulse (BVP), and galvanic skin response (GSR/EDA) and behavioral modalities (facial expressions, and speech) in a unified arousal-valence 2D interface to track moment-to-moment emotional states. This architecture enables interpretable, user-specific analysis and supports applications in emotion education, neuroadaptive feedback, and interaction support for neurodiverse users. Two demonstration scenarios illustrate its application: (1) passive media viewing (2D or VR videos) reveals cortical and autonomic responses to affective content, and (2) semi-scripted conversations with a facilitator or virtual agent capture real-time facial and vocal expressions. These tasks enable controlled and naturalistic emotion monitoring, making the system well-suited for personalized feedback and neurodiversity-informed interaction design.

## 1 Introduction

Automatic emotion recognition remains a central problem in affective computing, with applications ranging from adaptive human–computer interaction to mental health monitoring and education. Traditionally, emotional states are inferred from observable behavioral signals such as facial expressions and vocal prosody, or from physiological signals such as brain activity and autonomic responses. However, each modality presents distinct methodological challenges. Behavioral expressions of emotion are often modulated by social norms, context, or individual traits, and may be atypical or suppressed in neurodiverse populations, including individuals with autism spectrum disorder (ASD) or alexithymia, who may experience intense affect without overt expression [1,2]. On the other hand, physiological signals such as EEG, ECG, and GSR offer a more covert and continuous channel for affect sensing, but are vulnerable to motion artifacts, environmental noise, and inter-individual variability.

To improve robustness and inclusivity, multimodal emotion recognition approaches aim to integrate both behavioral and physiological signals. However, many systems adopt early or late fusion strategies that produce a single affective prediction, thereby masking the unique contributions and nuances of each modality [3, 4]. In addition, few systems support real-time, interpretable visualization of emotion estimates across modalities. A critical open challenge lies in the variability of emotional expression across individuals: for some, facial expressions or vocal prosody may reliably reflect internal states, while for others—physiological cues like EEG or ECG may provide more consistent indicators [1,2,5]. Understanding these inter-individual differences in modality informativeness is essential for developing affective technologies that adapt to users' expressive profiles, support emotion awareness, and enable personalized interventions.

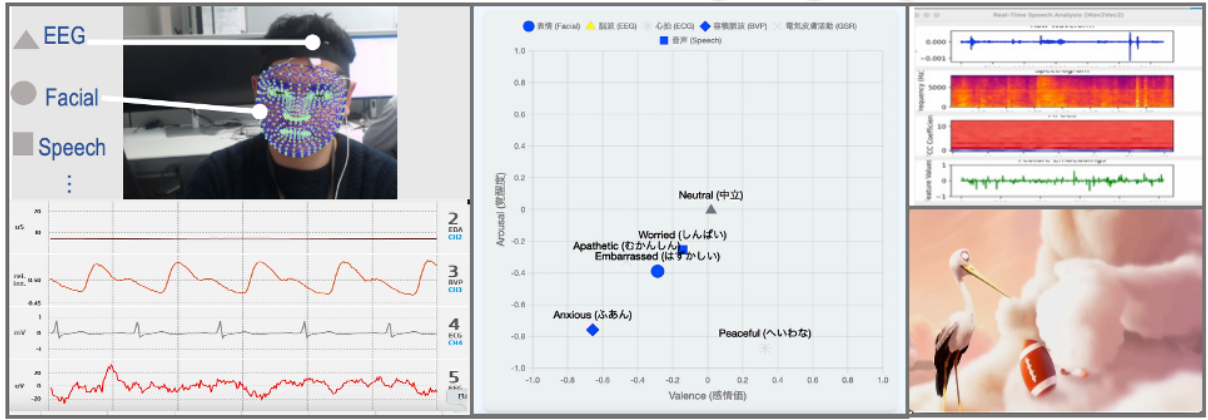

Figure 1: Main interface of the Multimodal Emotion Estimator displaying real-time predictions from facial expressions, speech, and neurophysiological signals (EEG, ECG, BVP, GSR) on a shared 2D arousal–valence map, alongside synchronized input and modality-specific GUIs (face mesh: upper left; speech features: upper right), and video stimulus (lower right).

In this demonstration, we present a real-time multimodal emotion estimation system that concurrently captures and visualizes arousal and valence estimates from behavioral (facial expressions, speech) and neurophysiological signals, including electroencephalography (EEG), electrocardiography (ECG), blood volume pulse (BVP), and galvanic skin response (GSR) or electrodermal activity (EDA). Rather than collapsing modalities into a fused prediction, each signal is independently projected into a shared two-dimensional space based on Russell's Circumplex Model of Emotion [6], which represents affective states along valence (unpleasant to pleasant) and arousal (calm to activated) axes, normalized between –1 and 1. This unified reference frame enables observation of alignment, divergence, and temporal dynamics across channels while allowing signal-specific analysis of intra-individual variation. The system supports future development of personalized, context-aware emotion sensing tools for neuroadaptive feedback, therapeutic use, and neurodivergent education.

## 2 The Multimodal Emotion Estimator System

Figure 1 shows that within the 2D space, each emotion estimate is rendered as a uniquely shaped point corresponding to its source modality—e.g., circle for facial expressions, square for speech, and triangle for EEG—allowing immediate visual differentiation across channels. The point's position is determined by the model's predicted arousal and valence values, while its color corresponds to the emotional quadrant it falls into: High Control (e.g., excitement, enthusiasm) is shown in yellow, Obstructive (e.g., anger, stress) in red, Low Control (e.g., sadness, boredom) in blue, and Conductive (e.g., calm, contentment) in green [7]. Point size encodes the intensity of the emotional response. Specifically, the Euclidean distance from the neutral center (0, 0) reflects the extremity of the predicted emotion, with larger points indicating greater intensity or emotional salience. Each prediction is also associated with a discrete emotion label. For physiological and speech signals, a valence–arousal-to-category mapping yields 34 interpretable labels (e.g., "Happy", "Frustrated") based on established dimensional-to-categorical conversion models [9, 10]. For facial expression data, the MorphCast Emotion AI SDK [9] provides an extended set of 38 emotion categories (Affects38), allowing fine-grained affect classification directly from webcam input. The system also provides synchronized time-series visualizations and graphical interfaces for each modality. When a specific channel (e.g., ECG) is selected, its waveform is displayed beneath the 2D arousal–valence emotion map; facial expressions are visualized with MediaPipe's face mesh overlaying 3D landmarks [8]; and speech is represented through waveform graphs and spectrograms [11].

Each modality is processed in real time through dedicated inference pipelines. Facial expressions are analyzed using the MorphCast Emotion AI SDK, a lightweight (<1 MB), offline browser-based model that enables fast, private, and local inference. Independent benchmarks on posed (BU-4DFE) and spontaneous (UT-Dallas) expression datasets report a 48.56% true positive rate in facial expression classification [11]. Speech emotion is estimated using the wav2vec2-large-robust dimensional model, which combines a frozen convolutional neural network (CNN) front-end with fine-tuned transformer layers to directly model raw audio. It achieves concordance correlation coefficients (CCC) of 0.745/0.636 (arousal/valence) on MSP-Podcast, 0.663/0.448 on IEMOCAP, and 0.539 on MOSI datasets [12]. EEG-based emotion recognition uses a 1D CNN trained on bandpower features (alpha, high/low beta, gamma, theta) from a single differential channel (Fp1–Fp2) using the DEAP dataset [13]. The ECG model adopts the WildECG backbone [14], pretrained on the TILES dataset and fine-tuned for valence–arousal regression using the YAAD dataset [15]. BVP and GSR models are trained on sliding-window DEAP data [13] using a two-stage framework: stage one uses PaPaGei for BVP [16] and a three-layer GRU for GSR [17], followed by a single-layer GRU on high-confidence segments in stage two. Each stage includes a two-layer MLP head. Although CCC scores for these physiological modalities remain below 0.5, ongoing efforts aim to improve accuracy through advanced architectures such as transformers and enriched features such as heart rate variability (HRV) dynamics and entropy-based descriptors. Future updates will also incorporate uncertainty-aware predictions to reflect confidence levels and support more interpretable multimodal visualization.

## 3  System Architecture and Signal Routing

The system is built around a central Python controller (main.py) that coordinates real-time execution of modular pipelines across multiple data streams. Neurophysiological signals are acquired using the BioSignalsPlux 8-channel hub [18], a Bluetooth-based, research-grade biosensor. The signals are sampled at 100 Hz (16-bit), visualized via OpenSignals, and streamed to the Python backend using the Lab Streaming Layer (LSL). Each stream undergoes modality-specific preprocessing—such as ADC-to-physical unit conversion, bandpass filtering, power spectral density computation, and normalization—before being passed to PyTorch or Keras models for valence and arousal inference. For EEG, our system supports BioSignalsPlux sensor and the FocusCalm headband [19], which transmits single-channel frontal EEG (Fp1–Fp2) via Wi-Fi using UDP packets and processed identically to LSL inputs. Facial expression analysis accesses webcam input while speech audio is captured using the built-in microphone. All data inputs are processed locally on a standalone laptop and is designed with strict ethical privacy safeguards: no data is stored, recorded, or transmitted externally during deployment, thereby mitigating risks associated with data leakage, remote access, or post-hoc misuse.

## 4  Demonstration tasks and setup

For neurophysiological signals, the demo presents emotion-eliciting video clips or VR simulations from validated affective media sources, shown on a 2D screen or VR headset to enhance immersion. For behavioral signals—facial expressions and speech—the demo involves semi-scripted conversations with a facilitator or virtual agent designed to evoke specific emotions. These two scenarios—passive stimulation and interactive dialogue—showcase the system's ability to monitor emotion dynamics across multiple modalities, supporting use cases in education, affective computing, and adaptive feedback.

## 5  Conclusions

This work presents a real-time multimodal emotion estimation system that independently models behavioral and neurophysiological signals within a unified arousal–valence space. By retaining modality-specific outputs, the system enables interpretable, context-aware visualization and supports intra-individual analysis—especially relevant for neurodivergent populations, where emotional expression may vary across channels. Demonstrations involving media viewing and conversational tasks highlight the system's capacity to capture dynamic emotional responses in structured and naturalistic settings while maintaining ethical integrity through fully local, storage-free processing. Ongoing improvements focus on refining preprocessing, enriching feature extraction, and adopting advanced architectures. Future updates will integrate subject-specific calibration mechanisms and uncertainty-aware prediction outputs, enhancing model robustness, interpretability, and inclusivity. Broader validation across diverse user populations remains a critical next step for ensuring real-world reliability and generalizability.

## Acknowledgments

This work was supported by the JST Moonshot R&D (Grant Number: JPMJMS2292), by the JSPS KAKENHI (Grant Number: 21H04981), by the International Research Center for Neurointelligence (IRCN) Catalyst and Pilot Research Support Grants, and by the World Premier International Research Center Initiative (WPI), MEXT, Japan.